\documentclass[RNAAS]{aastex63}

\usepackage{color}

\submitjournal{RNAAS}

\begin{document}

\title{
Triaxiality can explain the alleged dark matter deficiency in some dwarf galaxies}

\correspondingauthor{J. S\'anchez Almeida}
\email{jos@iac.es}

\author[0000-0003-1123-6003]{J. S\'anchez Almeida} 
\affil{Instituto de Astrof\'\i sica de Canarias, La Laguna, Spain} 
\affil{Universidad de La Laguna}

\author[0000-0001-7958-4536]{M.~Filho}
\affil{Faculty of Engineering, University Oporto, Oporto, Portugal}
\affil{CENTRA/SIM, Faculty of Sciences, University Lisbon, Lisbon, Portugal}

\section{Context} \label{sec:intro}

Dark Matter (DM) is an ingredient essential to the current cosmological concordance model ($\Lambda$CDM). It provides the gravitational pull needed for the baryons to form galaxies. Therefore, the existence of galaxies without DM is both disquieting and extremely interesting.\footnote{Although their existence can be accommodated without twisting the $\Lambda$CDM  paradigm; e.g., \citeauthor{2019A&A...626A..47H}~(\citeyear{2019A&A...626A..47H}).} 
  Thus, the finding of galaxies lacking DM  prompted much discussion in the technical literature \citep[e.g.,][]{2018Natur.555..629V,2019MNRAS.486.1192T}.


Recently, \citet{2019NatAs.tmp..493G} presented {\em further evidence for a population of dark-matter-deficient dwarf galaxies.} They found 19 dwarfs  that may consist only of baryons. They were selected from a sample of 324 galaxies in the ALFALFA HI survey \citep{2011AJ....142..170H} that have optical counterparts and so their baryon mass (stars plus gas; $M_{bar}$) can be measured. The total mass of the galaxies (i.e., DM plus baryons) was inferred from the HI line width after correcting for the line-of-sight inclination  ($i$) of the assumed HI disk. The unknown $i$ was estimated from the axial ratio $b/a$ measured on the optical images, leading to
\begin{equation}
M_{dyn}=M^i_{dyn}\,\sin^2i~\frac{1-(b/a)_0^2}{1-(b/a)^2},
\label{eq:1}
\end{equation}
where $M^i_{dyn}$ is the true dynamical mass, and $M_{dyn}$ represents the dynamical mass estimated from the minor ($b$) and major ($a$) axes of the galaxies projected in the plane of the sky. The thickness of the disk, $(b/a)_0$, was set to $0.2$.  \citet{2019NatAs.tmp..493G}  consider and discard various systematic effects associated with the misalignment between the HI and optical disks that may cause  $M_{dyn}\not= M^i_{dyn}$. However, they  bypass the triaxiality of the dwarf galaxies,
i.e., the fact that dwarfs are 3D objects with different sizes in the three axes, while for $M_{dyn} \simeq  M^i_{dyn}$ in Eq.~(\ref{eq:1})  galaxies have to be axisymmetric structures. Such oversimplification may cause
$M_{dyn} \ll M^i_{dyn}$ thus weakening their case for the existence of DM-deficient dwarfs.

\section{Effect of triaxiality on the dynamical mass estimate}

The galaxies are assumed to be disk-like so that when viewed almost face-on ($\sin i \rightarrow 0$) then $b/a\rightarrow 1$ and  Eq.~(\ref{eq:1}) gives $M_{dyn}\simeq M^i_{dyn}$. However, if the real galaxies are triaxial then $b/a\not= 1$ even when $\sin i\rightarrow 0$, leading to $M_{dyn} \ll M^i_{dyn}$ for small $i$. This effect was not considered by \citet{2019NatAs.tmp..493G} in their analysis, despite triaxiality being the rule among dwarf galaxies \citep[e.g.,][]{2013MNRAS.436L.104R,2019ApJ...883...10P}. The issue is whether triaxiality explains the apparent deficit of DM shown by some of their galaxies. 

\begin{figure}
\centering 
\includegraphics[width=0.8\textwidth]{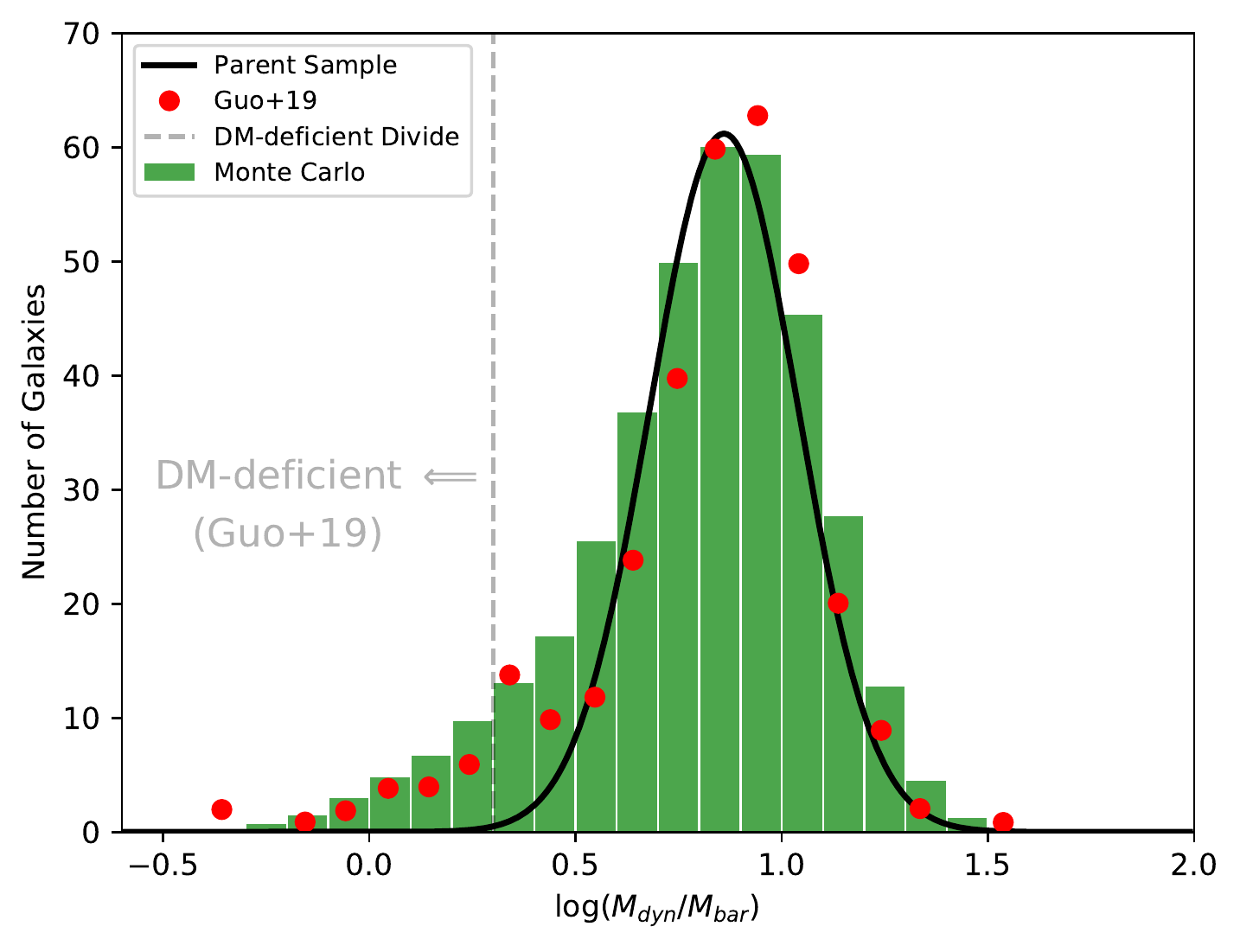}
\caption{Monte Carlo simulated distribution of $\log(M_{dyn}/M_{bar})$ resulting from the triaxiality-induced bias.
From a parent distribution with  $M^i_{dyn} >> M_{bar}$ (black line; Gaussian with mean and  standard deviation taken from the  parent sample of \citeauthor{2019NatAs.tmp..493G}~\citeyear{2019NatAs.tmp..493G}), the bias creates a long tail toward $M_{dyn} \simeq M_{bar}$ (green bars). The simulated  distribution closely resembles the distribution  inferred by \citet[][red symbols]{2019NatAs.tmp..493G}.  Almost face-on triaxial galaxies are never recognized as such in optical images, which systematically underestimate $M_{dyn}$ (Eq.~[\ref{eq:1}] with $b/a\not= 1$ even when $i\rightarrow 0$). All model galaxies have the same  baryon mass and 3D axial ratio ($A:B:C = 1:0.62:0.3$),  with random orientations independently of their true dynamical mass.  Following \citet{2019NatAs.tmp..493G},  only galaxies with $0.3 \le b/a \le 0.6$ are included in the simulated histogram, which is scaled to match the number of observed galaxies.}
\label{fig:nodm_fake} 
\end{figure}

A Monte Carlo simulation was designed  to address the impact of this triaxiality-induced bias on $M_{dyn}$. Specifically, we model the distribution of $M_{dyn}$ provided by Eq.~(\ref{eq:1}) from a non-DM-deficient distribution of $M^i_{dyn}$.  All model galaxies have the same 3D ellipsoidal shape set by the three semi-axes $A$, $B$, and $C$.\footnote{$A\ge B\ge C$,  $A=B \gg C$ for disks, and $A\not= B\not= C$ for triaxial objects.} The projection in the plane of the sky depends on the inclination and azimuth of the galaxy and can be computed analytically to get $b/a$ \citep[e.g.,][]{1998NCimB.113..927S}. Thus,  $M_{dyn}$ is set given $M^i_{dyn}$ and the 3D shape and orientation of the model galaxy \citep[see][]{2019ApJ...883...10P}. Assuming the orientation of  the model galaxies to be random and independent of $M^i_{dyn}$, one recovers a histogram for $\log(M_{dyn}/M_{bar})$ with an artificial tail toward  $\log(M_{dyn}/M_{bar})\sim 0$ (Fig.~\ref{fig:nodm_fake}, green bars), which is absent in the distribution of the true $\log(M^i_{dyn}/M_{bar})$ (Fig.~\ref{fig:nodm_fake}, black line). The model distribution  closely resembles the distribution inferred by \citet[][red symbols]{2019NatAs.tmp..493G}. The tail in the model is almost exclusively determined by $B/A$, set to 0.62,  a value consistent with the range found in literature \citep[e.g.,][]{2013MNRAS.436L.104R,2019ApJ...883...10P}.

Another independent argument also suggests bias. If unbiased, galaxies should have random orientations and, therefore, a uniform distribution of $b/a$ provided they are disks. If biased, the alleged DM-deficient galaxies should preferentially have large $b/a$ since they are almost face-on. The latter trend is present in the $b/a$ observed by \citet{2019NatAs.tmp..493G}. A Kolmogorov-Smirnoff test shows  with 92\,\%\ confidence that their $b/a$  are inconsistent with a uniform distribution, invalidating the use of $b/a$ to determine $i$.   
%

\section{Conclusions}

Our simulation shows how the triaxiality of dwarf galaxies must be considered to measure dynamical masses, calling into question that \citet{2019NatAs.tmp..493G} found  further evidence for a population of DM-deficient dwarf galaxies. Such a population may consist of normal almost face-on HI disks with their inclination overestimated.     



\end{document}